\documentclass{svjour3}                     
\smartqed  
\usepackage{graphicx}
\usepackage{amsmath}
 \usepackage{mathptmx}      
\begin{document}

\title{Neutrino masses beyond the tree level\thanks{Presented at the
    workshop ``30 years of strong interactions'', Spa, Belgium, 6-8
    April 2011.}}


\author{D. Aristizabal Sierra
}


\institute{
  \email{daristizabal@ulg.ac.be}  \\
  \emph{Present address:} IFPA, Dep. AGO, Universite de Liege, Bat
  B5 \small \sl Sart Tilman B-4000 Liege 1, Belgium
}

\date{
}

\maketitle

\begin{abstract}
  Models for Majorana neutrino masses can be classified according to
  the level in perturbation theory at which the effective dimension
  five operator $LLHH$ is realized. The possibilities range from the
  tree-level up to the three-loop level realizations. We discuss some
  general aspects of this approach and speculate about a model
  independent classification of the possible cases. Among all the
  realizations, those in which the effective operator is induced by
  radiative corrections open the possibility for lepton number
  violation near -or at- the electroweak scale. We discuss some
  phenomenological aspects of two generic radiative realizations: the
  Babu-Zee model and supersymmetric models with bilinear R-parity
  violation.  
\end{abstract}

\section{Introduction}
\label{sec:intro}
Neutrino experiments have firmly demonstrated neutrinos are massive
and have non-vanishing mixing angles among the different generations
\cite{Schwetz:2008er}. Since in the standard model neutrinos are
massless this experimental results are a clear evidence of beyond
standard model physics. From a general point of view Majorana neutrino
masses can be generated by adding to the standard model Lagrangian the
non-renormalizable effective Lagrangian \cite{Weinberg:1980bf}
\begin{equation}
  \label{eq:eff-op}
  {\cal L}_5\sim\frac{1}{\Lambda_{\text{NP}}} {\cal O}_5=
  \frac{1}{\Lambda_{\text{NP}}} 
  \left(\overline{L^c}i\tau_2H\right)\; 
  \left(H i\tau_2L\right)\,,
\end{equation}
where $L$ and $H$ are the lepton and Higgs electroweak doublets.  The
presence of the effective Lagrangian in (\ref{eq:eff-op}) guarantees
that after electroweak symmetry breaking neutrinos acquire Majorana
masses. Extensions of the standard model in which neutrinos have
Majorana masses realize the effective operator ${\cal O}_5$ in
different ways. However, among all these possibilities there are
subsets that have a common feature, namely the order in perturbation
theory at which the operator is realized. As will be discussed in
sec.~\ref{sec:O5realizations} a rather general classification of
neutrino mass models can be done by gathering together these different
subsets.

Making an exhaustive list of possible realizations of ${\cal O}_5$ and
their corresponding phenomenology is far out of the scope of the
present discussion. Thus, instead of using this approach we will
discuss two classes of generic models: models in which neutrino masses
are generated radiatively at the 2-loop level and supersymmetric
models with broken R-parity. In the former case we will stick to the
Babu-Zee model \cite{Zee:1985id,Babu:1988ki}, while for the latter we
will discuss bilinear R-parity breaking models with a neutralino LSP
\cite{Porod:2000hv}.
\section{The ${\cal O}_5$ operator and its different realizations}
\label{sec:O5realizations}
At $\Lambda_{\text{NP}}$ the heavy physical degrees of freedom, that
when integrated out yield the dimension five ($d=5$) effective
operator ${\cal O}_5$, are no longer decoupled. Once specified they
define a particular model for Majorana neutrino mass generation, so
different models lead to different realizations of ${\cal O}_5$. As
pointed out in the introduction these realizations can be classified
according to the order in perturbation theory at which ${\cal O}_5$ is
generated, namely tree-level (${\cal O}_5^{\ell=0}$), one-loop (${\cal
  O}_5^{\ell=1}$), two-loops (${\cal O}_5^{\ell=2}$) and three-loops
(${\cal O}_5^{\ell=3}$), where $\ell$ denotes the number of loops in
each case.  Three-loop level realizations require order one Yukawa
couplings implying that models based on ${\cal O}_5^{\ell\geq 4}$ are,
in general, not consistent with neutrino data once the requirement of
perturbativity of the corresponding couplings is imposed.

A general and model independent classification of the tree-level
${\cal O}_5^{\ell=0}$ and one-loop level ${\cal O}_5^{\ell=1}$
realizations of ${\cal O}_5$ has been carried out in
ref. \cite{Ma:1998dn} and such an analysis could be, in principle,
extended to the two and three loop order realizations. Regardless of
the order at which the operator arises the procedure is based on the
determination of all the possible gauge invariant renormalizable
vertices within the loop for all the different possible topologies. In
this procedure the only gauge quantum numbers that are fixed are those
of the external legs ($LLHH$) while those of the physical degrees of
freedom flowing in the loop are free and should be fixed by the
requirement of gauge invariance. Another approach for the same sort of
classification is by using all the possible $\Delta L=2$ effective
operators up to certain $d$. This method has been implemented in
ref. \cite{Babu:2001ex} for effective operators up to $d=11$. The
difference between these two approaches is that while the former focus
exclusively on the ${\cal O}_5$ operator the later covers this
operator but also higher dimensional operators as e.g. ${\cal O}_7\sim
(LLHH)(H^\dagger H)$. Note that these higher dimensional effective
operators ($d>5$) can give a dominant contribution to neutrino masses
only if the leading effective operator ${\cal O}_5$ is forbiden due to
a new symmetry. These cases have been throughout analysed in
ref. \cite{Bonnet:2009ej} and we have nothing more to add here.

The tree-level realizations of ${\cal O}_5$ correspond to the
different type of seesaw models (type-I \cite{seesaw}, type-II
\cite{Schechter:1980gr} and type-III \cite{Foot:1988aq}) whereas for
the 1-loop cases the number of possibilities (models) is much more
larger (is determined by the different $SU(3)\times SU(2)\times U(1)$
assignments of the internal degrees of freedom flowing in the
loop). Examples of ${\cal O}_5^{\ell=1}$ include the Zee model
\cite{Zee:1980ai}, models with scalar leptoquarks
\cite{AristizabalSierra:2007nf} and models with extra scalars and
fermions with nontrivial color charges
\cite{FileviezPerez:2009ud}. Examples of two-loop realizations include
extended scalar sectors as in the case of the the Babu-Zee model
\cite{Zee:1985id,Babu:1988ki} and models with scalar leptoquarks
\cite{Babu:2001ex,Babu:2010vp}. Up to our knowledge three-loop level
realizations rely on extensions of both the scalar and fermion
sectors, examples can be found in references
\cite{Babu:2001ex,Krauss:2002px}.

In models in which the operator ${\cal O}_5$ is generated beyond the
tree-level the lepton number breaking scale can readily be at or
around the electroweak scale. Thus, models embedded in such
realizations usually lead to testable predictions in either
high-energy or high-intensity experiments. In what follows we will
discuss two main cases, the Babu-Zee model and supersymmetric bilinear
R-parity violating models, paying special attention to some of their
phenomenological implications.
\section{Two loop realization: the Babu-Zee model}
\label{sec:two-loop}
In this model the standard model scalar sector is extended by the
addition of two new scalars, $h^+$ and $k^{++}$, both singlets under
$SU(2)$. Their couplings to standard model leptons is given by
\begin{equation}\label{yuks}
{\cal L} = f_{\alpha\beta} (L^{Ti}_{\alpha L}CL^{j}_{\beta L})\epsilon_{ij}h^+
         + h'_{\alpha\beta}(e^T_{\alpha R}Ce_{\beta R})k^{++} + {\rm h.c.}
\end{equation}
Here, $L_L$ are the standard model (left-handed) lepton doublets, $e_R$ 
the charged lepton singlets, $\alpha ,\beta$ are generation indices and 
$\epsilon_{ij}$ is the completely antisymmetric tensor. Note that $f$ 
is antisymmetric, while $h'$ is symmetric. Assigning $L=2$ to $h^-$ and 
$k^{++}$, eq.  (\ref{yuks}) conserves lepton number. Lepton number violation 
in the model resides only in the following term in the scalar potential
\begin{equation}\label{scalar}
{\cal L} = - \mu h^+h^+k^{--} + {\rm h.c.}
\end{equation}
Here, $\mu$ is a parameter with dimension of mass.

\begin{figure}
  \begin{center}
    \includegraphics[width=60mm, height=30mm]{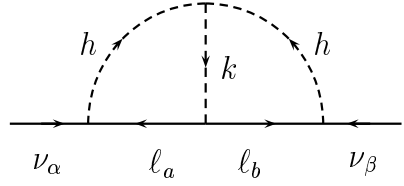}
  \end{center}
  \caption{Feynman diagram for the 2-loop Majorana neutrino masses in 
    the model of \cite{Zee:1985id,Babu:1988ki}.}
  \label{fig:Mnu}
\end{figure}

The setup of eq. (\ref{yuks}) and eq. (\ref{scalar}) generates Majorana 
neutrino masses via the two-loop diagram shown in fig. (\ref{fig:Mnu}). 
The resulting neutrino mass matrix can be expressed as 
\begin{equation}\label{mnu}
{\cal M}^{\nu}_{\alpha\beta} = \frac{8\mu}{(16 \pi^2)^2 m_h^2}
f_{\alpha x}\omega_{xy}f_{y\beta}{\cal I}(\frac{m_{k}^2}{m_{h}^2}), 
\end{equation}
with summation over $x,y$ implied. The parameters $\omega_{xy}$ are defined 
as $\omega_{xy}= m_x h_{xy} m_y$, with $m_x$ the mass of the charged lepton 
$l_x$. Following \cite{Babu:2002uu} we have rewritten $h_{\alpha\alpha}= 
h'_{\alpha\alpha}$ and $h_{\alpha\beta}=2 h'_{\alpha\beta}$. ${\cal I}(r)$ 
finally is a dimensionless two-loop integral given by
\begin{equation}\label{itilde}
{\cal I}(r) = - \int_0^1 dx \int_0^{1-x} dy 
\frac{1}{x+(r-1)y+y^2}\log\frac{y(1-y)}{x+ry} .
\end{equation}
For non-zero values of $r$, ${\cal I}(r)$ can be solved only numerically. 
We note that for the range of interest, say $10^{-2} \le r \le 10^{2}$, 
${\cal I}(r)$ varies quite smoothly between (roughly) $3 \le {\cal I}(r) \le 
0.2$.

\subsection{Flavour violating charged lepton decays}
\label{sec:lfv}
\begin{figure}
\centering
\includegraphics[width=5.6cm, height=5cm]{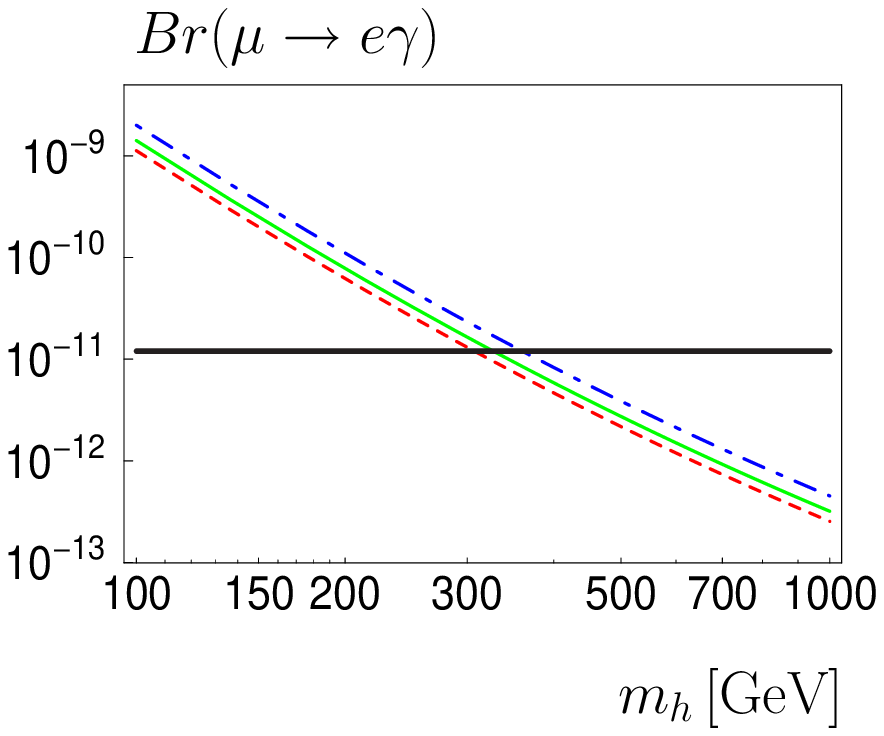}
\hspace{0.2cm}
\includegraphics[width=5.6cm, height=5cm]{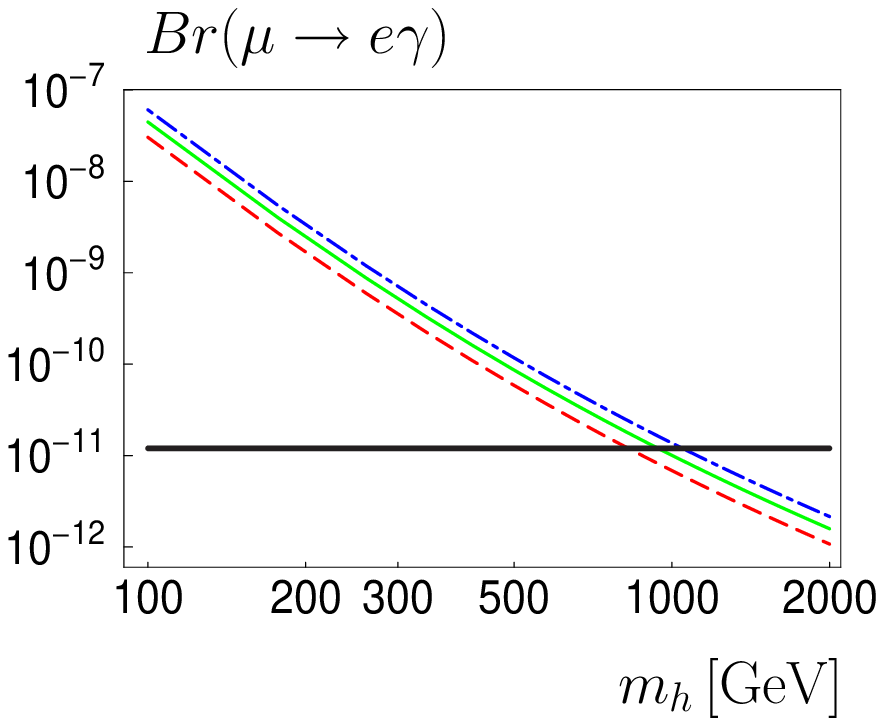}
\caption{Conservative lower limit on the branching ratio 
Br($\mu\rightarrow e\gamma$) as a function of the charged scalar 
mass $m_h$ for normal hierarchy (left plot) and inverted hierarachy (right plot). 
The three lines are for the current 
solar angle $\sin^2\theta_{12}$ best fit value (full line) and 3 $\sigma$ 
lower (dashed line) and upper (dot-dashed line) bounds. Other parameters fixed at 
$\sin^2\theta_{23}=0.5$, $\sin^2\theta_{13}=0.040$ and 
$\Delta m^2_{\rm Atm} = 2.0 \cdot 10^{-3}$ eV$^2$.}
\label{MuEGamVmH}
\end{figure}

Phenomenological tests of this model have been studied in
\cite{Babu:2002uu,AristizabalSierra:2006gb,Nebot:2007bc}.  Among all
of them those involving $\mu\to e\gamma$ can be regarded as the most
stringent ones. In ref. \cite{AristizabalSierra:2006gb} it has been
shown that the corresponding decay branching ratio for this process
can be written as
\begin{eqnarray}\label{eq:muegam}
\label{eq:muegam_num}
Br(\mu \rightarrow e\gamma)& \simeq & 4.5 \cdot 10^{-10} 
\Big(\frac{\epsilon^2}{h_{\mu\mu}^2{\cal I}(r)^2}\Big) 
\Big(\frac{m_{\nu}}{\rm 0.05 \hskip1mm eV}\Big)^2
\Big(\frac{\rm 100 \hskip1mm GeV}{m_{h}}\Big)^2\,,
\end{eqnarray}
with $\epsilon=f_{e\tau}/f_{\mu\tau}$ and $m_h$ the mass of the singly
charged scalar. Figure \ref{MuEGamVmH} shows the resulting lower limit
on $Br(\mu\to e\gamma$) as a function of $m_h$ for the case of normal
and inverted hierarchies. Note that the horizontal solid line
indicates the upper limit set by the MEGA experiment
\cite{Brooks:1999pu} and not the new one placed by the MEG experiment,
$Br(\mu\to e \gamma)<2.4\times 10^{-12}$ at 90\%
C.L. \cite{Adam:2011ch}. Using the updated limits the constraints on
the singly charged scalar mass would be even more stringent that the
ones quoted here.

In summary, in this model $Br(\mu\to e\gamma)\geq 10^{-13}$ is
guaranteed for singly charged scalar masses smaller than 590 GeV (5.04
TeV) for normal (inverse) hierarchical neutrino masses, and larger or
even much larger branching ratios are expected in general.  Thus, a
non-observation of this process in the next few years, at least for
the case of inverse hierarchy, would certainly remove most of the
motivation to study this model.
\section{Bilinear R-parity violating supersymmetry}
\label{sec:r-parity-breaking}
Bilinear R-parity violation (BRpV) provides an intrinsically
supersymmetric framework for Majorana neutrino masses (for a review
see \cite{Hirsch:2004he}). In these models the superpotential includes, in
addition to the MSSM terms, also the term
\begin{equation}
  \label{eq:superpotentialBRpV}
  W_{\text{BRpV}}=\epsilon_i\hat L_i \hat H_u\,.
\end{equation}
This term breaks not only R-parity but also lepton number in all three
generations. In order to have a consistent model a soft SUSY breaking
term has to be added to the scalar potential, namely
\begin{equation}
  \label{eq:softV}
  V_{\text{BRpV}}^{\text{soft}}=B_i\epsilon_i\tilde L_i H_u\,.
\end{equation}
The presence of these terms induce a non-vanishing vacuum expectation
value for the sneutrinos ($v_i=\langle\tilde\nu_i\rangle$) that give
rise to a mixing among neutral gauginos and Higgsinos with neutrinos.
Due to this mixing one of the neutrinos acquires mass.
The effective neutrino mass matrix reads
\begin{equation}
  \label{eq:tree-level-mass}
  (m_\nu^{(0)})_{ij}=\frac{M_1 g^2 + M_2g'^2}
  {4\text{det}\,(\pmb{M_{\chi^0}})}\;\Lambda_i\Lambda_j\,,
\end{equation}
where $\Lambda_i=\mu v_i + v_d\epsilon$ (with $\mu$ the Higgsino mass
term and $v_d=\langle H_d \rangle$) and $\pmb{M_{\chi^0}}$ is the
neutralino mass matrix. The other two neutrinos acquire mass from
one-loop corrections involving $W$, $Z$ and scalar loops, being the
bottom-sbottom and tau-stau loops the most important contributions
\cite{Hirsch:2000ef}. Thus, the BRpV model is an example of a model in
which neutrino masses arise from different realizations of the ${\cal
  O}_5$ operator.

The atmospheric and reactor angles are approximately given by the rotation
angles that diagonalize the tree level mass matrix $m_\nu^{(0)}$:
\begin{equation}
  \label{eq:Atm-Reac-mixing-angles}
  \tan^2\theta_{23}\approx\frac{\Lambda_2^2}{\Lambda_3^2},\quad
  \tan^2\theta_{13}\approx\frac{\Lambda_1^2}{\Lambda_3^2+\Lambda_2^2}\,.
\end{equation}
The solar angle instead is obtained once the one-loop corrections are
taken into account. A remarkable feature of the bilinear R-parity
breaking model is that the same parameters that determine neutrino
physics also control the decay patterns of the LSP, thus it is always
possible to establish a set of correlations between the LSP decay
branching ratios and neutrino observables. These correlations can be
used as a tool to know whether BRpV is responsible for the origin of
neutrino masses.
\subsection{LSP decays and neutrino observables}
\label{sec:LSPdecays}
Once R-parity is broken the LSP is unstable and decays to standard
model fermions.  Thus astrophysical contraints on its nature do not
hold any more and in principle any supersymmetric particle can be the
LSP.  The decay patterns of all possible LSPs and their relations with
neutrino observables in the context of bilinear R-parity breaking
models have been analysed in
\cite{Porod:2000hv,Hirsch:2003fe}. Additional studies in more generic
models including trilinear R-parity violating couplings have been
carried out for slepton and sneutrino LSPs
\cite{Bartl:2003uq,Aristizabal Sierra:2004cy}. Here we will highlight
some of the main features of this ``program'' in bilinear R-parity
breaking models assuming the lightest neutralino to be the LSP. The
following discussion is based on ref. \cite{Porod:2000hv}.

The presence of the bilinear R-parity violating parameters induce not
only a mixing between neutralinos and neutrinos but also a mixing
between charginos and charged leptons, charged Higgses and sleptons,
CP-even (CP-odd) components of the neutral Higgses and the
corresponding CP-even (CP-odd) components of the sneutrinos
\cite{Hirsch:2000ef}. All together these mixings determine the
three-body leptonic, semi-leptonic and invisible neutralino decays:
$\tilde\chi^0_1\to \nu_i l_j^+ l_k^-$, $\tilde\chi^0_1\to l^\pm_iq\bar
q'$, $\tilde\chi^0_1\to \nu_i q\bar q$ and $\tilde\chi^0_1\to
\nu_i\nu_j\nu_k$.

\begin{figure}
  \centering
  \includegraphics[width=6cm,height=4.6cm]{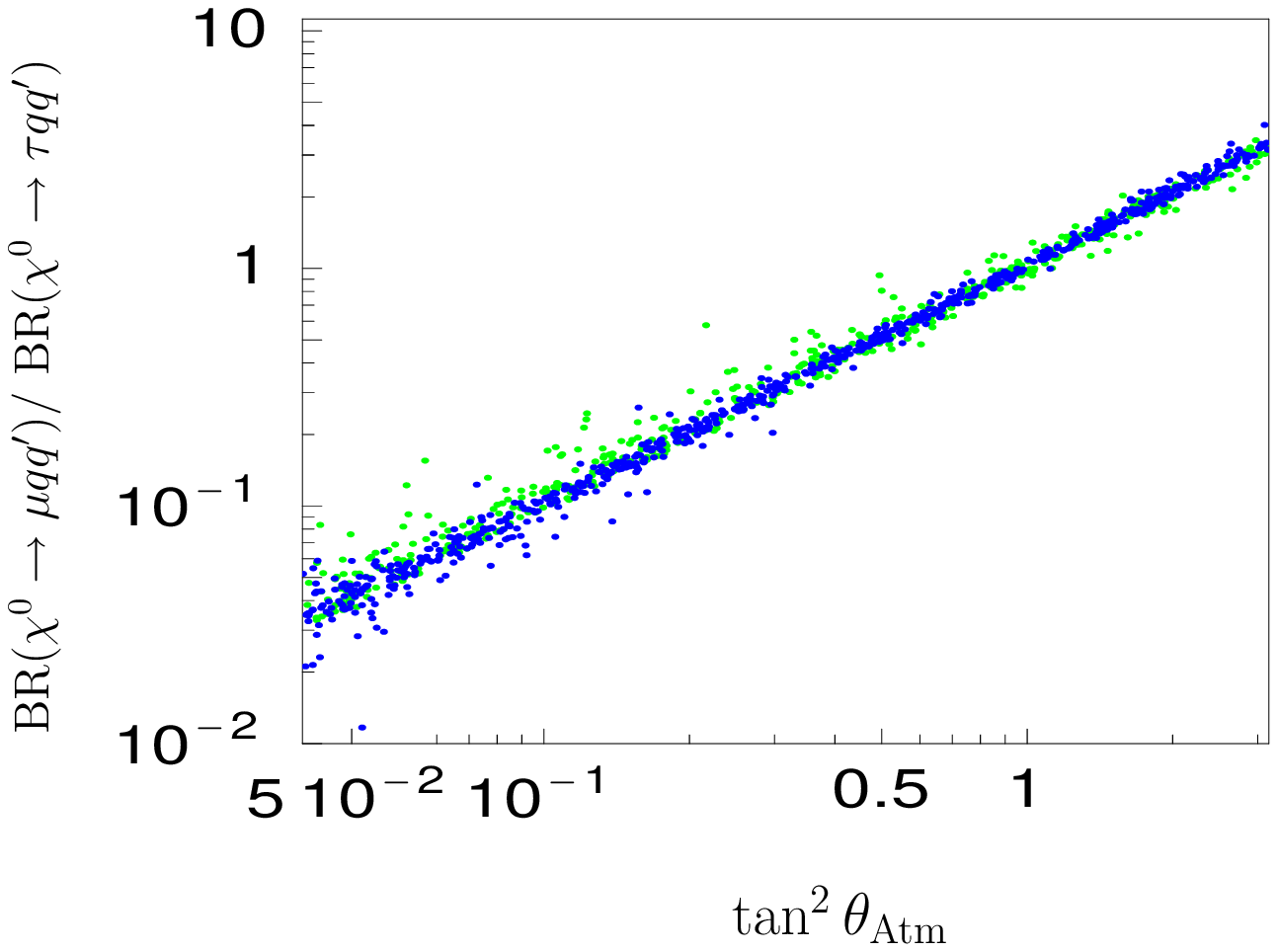}
  \includegraphics[width=6cm,height=4.6cm]{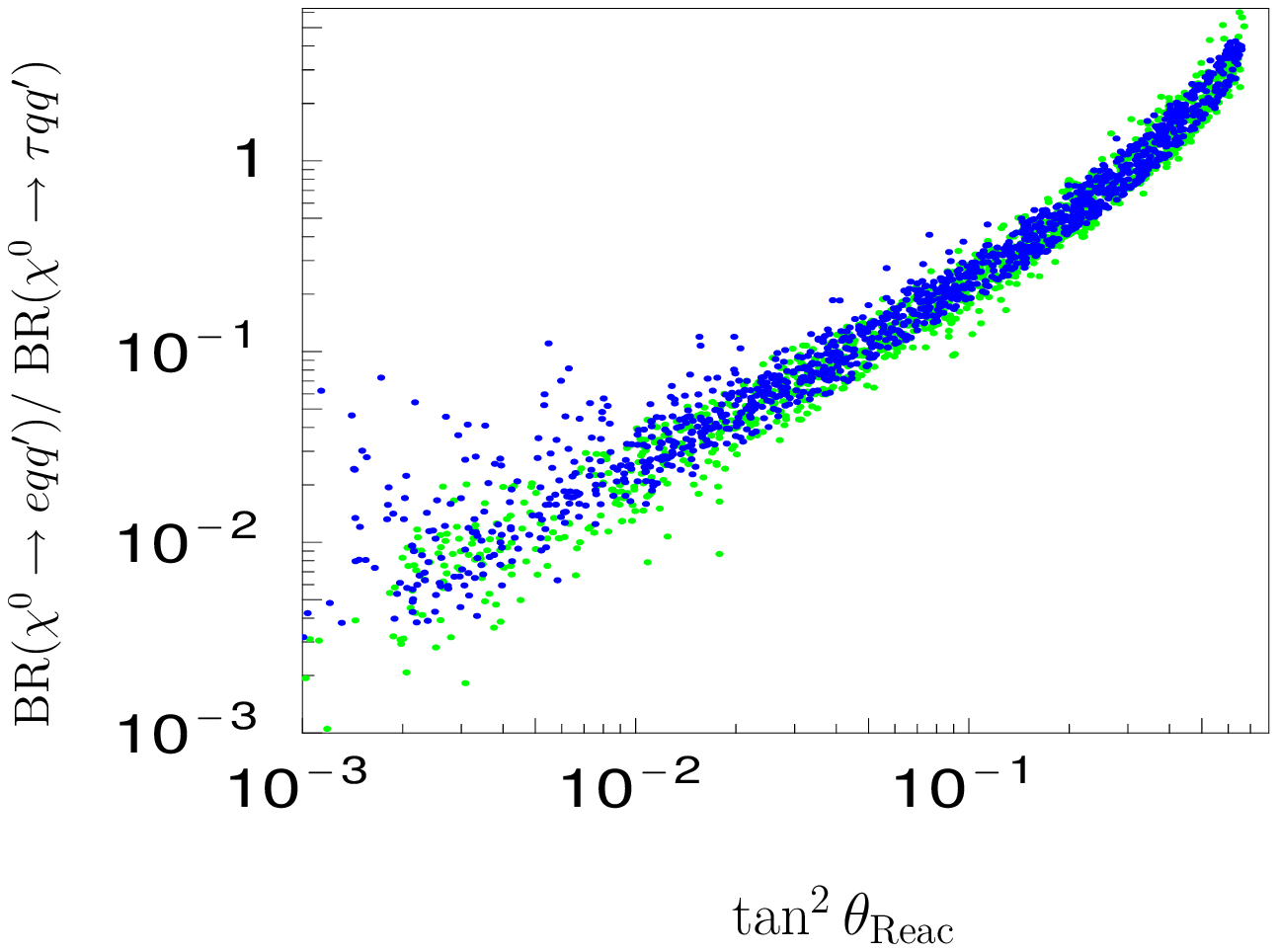}
  \caption{Ratios of decay branching ratios for semi-leptonic
    neutralino decays and their correlations with the atmospheric and
    reactor angles \cite{Porod:2000hv}.}
  \label{fig:neutralino-decayBR}
\end{figure}

Given the above discussion it should now become clear that once the
bilinear R-parity breaking parameters are constrained by neutrino
experimental data the lightetest neutralino decays are expected to be
constrained as well. In fact, it turns out that the constraints
imposed by neutrino physics assure: ($i$) the neutralino always decay
inside the detector; ($ii$) the decay branching ratio for neutralino
invisible decays never exceeds 10\%; ($iii$) different ratios of decay
branching ratios are strongly correlated with neutrino mixing angles.
Figure \ref{fig:neutralino-decayBR} shows the corresponding
correlations for the semi-leptonic final states $\mu qq'$, $\tau qq'$
and $e qq'$. From these results and the measured neutrino mixing
angles it can be established that if BRpV is the mechanism responsible
for the origin of neutrino masses and the lightest neutralino turn out
the be the LSP the following measurements should be expected at LHC:
\begin{equation}
  \label{eq:BRrelations}
  \text{BR}(\tilde \chi^0_1\to \mu qq')\simeq\text{BR}(\tilde \chi^0_1\to \tau qq')\quad
  \text{and}\quad
  \text{BR}(\tilde \chi^0_1\to e qq')\ll\text{BR}(\tilde \chi^0_1\to \mu qq')\,.
\end{equation}

In summary, in bilinear R-parity breaking models the decay patterns of
the LSP are strongly correlated with neutrino mixing angles. These
correlations allow to set constraints on the different decay branching
ratios of the LSP that in turn can be used to prove whether these
models are responsible for the origin of neutrino masses and mixings.
\section{Conclusions}
\label{sec:conclusions}
From a general perspective Majorana neutrino masses can be accounted
for by the effective dimension five operator ${\cal O}_5$. We have
argued that the different ``incarnations'' of this operator can be
classified according to the order in perturbation theory at which the
operator is realized. In principle by using group theoretical
arguments one could make, in a model independent way, an exhaustive list
of all the possibilities at each order and up to the three-loop level
\footnote{Four-loop level realizations are in general incompatible
  with the measured neutrino mass scales once the requirement of
  perturbativity of the couplings is imposed.}.

Models in which ${\cal O}_5$ is realized radiatively (${\cal
  O}_5^{\ell\neq 0}$, where $\ell$ denotes the numer of loops) rely on TeV
scale physics. Thus, an obvious question is whether this new physics,
and therefore the origin of neutrino masses, can be proved at e.g. the
LHC. As illustrative examples we have discussed what we consider two
benchmark models: The Babu-Zee model and supersymmetry with bilinear
broken R-parity.
\begin{acknowledgements}
\label{sec:acnowledgements}
I would like to thank Sergey Kovalenko, Werner Porod and Diego
Restrepo for the enjoyable collaboration on some of the subjects
discussed here. I want to especially thank Martin Hirsch for the
collaboration that led to some of the papers quoted here and for the
always enlightening discussions.
\end{acknowledgements}

\end{document}